\documentclass[aps,showpacs,preprintnumbers,amsmath,amssymb,amsfonts,superscriptaddress,openany,nofootinbib,notitlepage]{revtex4-1}
\usepackage[utf8]{inputenc}
\usepackage{amsfonts}
\usepackage{amsbsy}
\usepackage{url}
\usepackage{enumerate}
\usepackage[justification=raggedright]{caption}
\usepackage{bbm}

\usepackage{graphicx,color}
\usepackage{amssymb,amsmath}
\usepackage{slashed}
\usepackage{hyperref}
\usepackage{braket}
\usepackage{simplewick}
\usepackage[title]{appendix}
\usepackage{caption}

\usepackage{ascmac}
\usepackage{latexsym}
\usepackage{pifont}          
\usepackage{bm}

\newcommand{\n}{\nonumber}

\newcommand{\mr}[1]{\mathrm{#1}}

\newcommand{\h}[1]{\hspace{#1}}
\newcommand{\f}[2]{\frac{#1}{#2}}

\begin{document}

\preprint{YGHP-19-01, KUNS-2759}

\title{Topological Nambu monopole in two Higgs doublet models}

\author{Minoru~Eto}
\affiliation{Department of Physics, Yamagata University, Kojirakawa-machi 1-4-12, Yamagata, Yamagata 990-8560, Japan}
\affiliation{Research and Education Center for Natural Sciences, Keio University, 4-1-1 Hiyoshi, Yokohama, Kanagawa 223-8521, Japan}

\author{Yu~Hamada}
\thanks{Corresponding author. \\E-mail address: yu.hamada@gauge.scphys.kyoto-u.ac.jp}
\affiliation{Department of Physics, Kyoto University, Kitashirakawa, Kyoto 606-8502, Japan}

\author{Masafumi~Kurachi}
\affiliation{Research and Education Center for Natural Sciences, Keio University, 4-1-1 Hiyoshi, Yokohama, Kanagawa 223-8521, Japan}

\author{Muneto~Nitta} 
\affiliation{Department of Physics, Keio University, 4-1-1 Hiyoshi, Kanagawa 223-8521, Japan}
\affiliation{Research and Education Center for Natural Sciences, Keio University, 4-1-1 Hiyoshi, Yokohama, Kanagawa 223-8521, Japan}

\begin{abstract}
We show that a topological Nambu monopole exists as a regular solution for a large range of parameters in two Higgs doublet models, 
contrary to the standard model admitting only non-topological Nambu monopoles.
We analyze a Higgs potential with a global $U(1)$ symmetry and a discrete symmetry $\mathbb{Z}_2$.
The monopole is attached by two topological $Z$ strings ($Z$ flux tubes) from both sides.
Despite of a trivial second homotopy group, the discrete symmetry $\mathbb{Z}_2$ together with a non-trivial first homotopy group for $Z$ strings topologically ensures the topological stability.
After analytically constructing an asymptotic form of such a configuration, we explicitly construct a solution of the equation of motion based on a 3D numerical simulation, in which magnetic fluxes spherically emanating from the monopole at large distances are deformed in the vicinity of the monopole.
Since the monopoles are expected to be abundant in the present universe, they might be observed in the current monopole searches.
\end{abstract}

\maketitle

\section{Introduction}
Magnetic monopoles have attracted great interests from both experimental and theoretical physicists since the seminal work by Dirac \cite{Dirac:1931kp}, 
which was motivated to improve the asymmetry between electric and magnetic charges in the Maxwell equations, providing an explanation for the electric charge quantization.
In field theoretical models, they have been theoretically realized as a regular solution \cite{tHooft:1974kcl,Polyakov:1974ek}
and have played crucial roles to study non-perturbative aspects of (non-)supersymmetric field theories \cite{Nambu:1974zg,Seiberg:1994aj,Seiberg:1994rs}.
However, except for condensed-matter analogues \cite{Castelnovo:2007qi,Ray:2014sga}, such monopoles have never been found in reality;
for instance, such magnetic monopoles are predicted by all grand unified theories (GUTs) \cite{Dokos:1979vu,Lazarides:1980va,Callan:1982au,Rubakov:1982fp}, 
and their search have been extensively conducted.
Nevertheless, no GUT monopoles have been found probably because the cosmological inflation diluted them.
To avoid dilution, monopoles should be produced below the inflation scale such as the electroweak scale.
In fact, a monopole configuration in the Standard Model (SM) was first considered by Nambu \cite{Nambu:1977ag}.
However, it is unstable because it must be attached from one direction by a string and the tension of the string pulls the monopole to infinity, although they were suggested to produce primordial magnetic fields before disappearance
\cite{Vachaspati:2001nb,Poltis:2010yu}.

In this paper, we predict an existence of a topologically stable
\footnote{
In this paper, the terminology ``stable'' is used for the stability of a single soliton put in the system.
In the context of cosmology, such a situation should be realized after the scaling regime is achieved 
as is known in the axion strings \cite{Sikivie:2006ni,Hagmann:1990mj,Shellard:1997mf}. 
} 
and magnetic monopole without singularity in a quite simple extension of the SM, two Higgs doublet model (2HDM), in which one more Higgs doublet is added to the SM (for reviews, see, e.g., Refs.\cite{Gunion:1989we,Branco:2011iw}).
They might be observed in the current monopole searches since they are expected to be abundant in the present universe.
Apart from cosmological production, our monopoles might also be produced and detected by the MoEDAL experiment \cite{Acharya:2019vtb} at LHC with masses of the order of a few TeV.
Their discovery would not only be a realization of Dirac's hypothesis but also yield a solid evidence of new physics beyond the SM, since there are no stable and regular magnetic monopoles in the SM.

The reason of the non-existence of stable monopoles in the SM is its trivial topology, that is,
the vacuum manifold is $S^3$ having a trivial second homotopy group $\pi_2$, as well as trivial $\pi_0$ for domain walls and $\pi_1$ for cosmic strings.
Nevertheless, (non-topological)
electroweak $Z$-strings (or magnetic $Z$-fluxes)
\cite{Vachaspati:1992fi,Vachaspati:1992jk,Achucarro:1999it,Brandenberger:1992ys,Barriola:1994ez}
have been studied, 
but they were shown to be unstable in realistic parameter region 
\cite{James:1992zp,James:1992wb}. 
Nambu monopoles are the end points of these strings \cite{Nambu:1977ag}.


2HDM is one of the most popular extensions of the SM with a potential to solve problems that are unanswered by the SM.
It has four additional scalar degrees of freedom in addition to 125 GeV Higgs boson ($h$), which are charged Higgs bosons ($H_{\pm}$), CP-even Higgs boson ($H$) and CP-odd Higgs boson ($A$). 
These additional scalars can be directly produced at LHC, though there is no signal so far today, placing lower bounds on masses of those additional scalar bosons.
Those lower bounds highly depend on parameter choices of 2HDM as well as how SM fermions couple to the two doublets.
For more detailed phenomenological studies, see, e.g., Refs. \cite{Trodden:1998ym,Kanemura:2015mxa,Kanemura:2014bqa,Kling:2016opi,Haller:2018nnx} and references therein.
Moreover, 2HDM has a much richer vacuum structure than the SM,
therefore allowing a variety of 
topologically stable solitons, 
in addition to non-topological solitons \cite{La:1993je,Earnshaw:1993yu,Perivolaropoulos:1993gg,Bimonte:1994qh,Ivanov:2007de,Brihaye:2004tz,Grant:2001at,Grant:1998ci,Bachas:1996ap} analogous to the SM;
domain walls \cite{Battye:2011jj,Brawn:2011,Eto:2018tnk,Eto:2018hhg}, 
membranes \cite{Bachas:1995ip,Riotto:1997dk}, 
and cosmic strings 
such as topological $Z$ strings \cite{Dvali:1993sg,Dvali:1994qf,Eto:2018tnk,Eto:2018hhg} (see also Ref. \cite{Bachas:1998bf}). 
However, magnetic monopoles were not examined because of a trivial second homotopy group $\pi_2$ as in the SM.
Instead, the stability of our monopole is topologically protected by a combination of the following two symmetries of the Lagrangian;
One is a global $U(1)$ symmetry that ensures the stability of the topological $Z$ strings.
The other is a discrete symmetry $\mathbb{Z}_2$ exchanging the topological strings.
Consequently, our monopole is attached by two topological $Z$ strings on {\it both sides}, where a $Z$ flux is confined on each string and the string tensions pulling the monopole are balanced due to the $\mathbb{Z}_2$ symmetry.
We explicitly construct such a solution of the equation of motion based on a 3D full numerical simulation, in which magnetic fluxes emanate from the monopole.
\\

\section{The model}
We introduce two $SU(2)$ doublets, $\Phi_1$ and $\Phi_2$, both with the hypercharge $Y=1$. The Lagrangian which describes the electroweak and the Higgs sectors is written as
\begin{align}
\h{-1em} {\mathcal L} = - \frac{1}{4}\left(Y_{\mu\nu}\right)^2 - \frac{1}{4}\left(W_{\mu\nu}^a\right)^2  + \left|D_\mu \Phi_i \right|^2 - V(\Phi_1, \Phi_2).
\end{align}
Here, $Y_{\mu\nu}$ and $W^a_{\mu\nu}$ describe field strength tensors of the hypercharge and the weak gauge interactions with $\mu$ ($\nu$) and $a$ being Lorentz and weak iso-spin indices, respectively. 
$D_\mu$ represents the covariant derivative acting on the Higgs fields, and the index $i$ runs $i=1,2$.
$V(\Phi_1, \Phi_2)$ is the potential for the two Higgs doublets.
In this paper, we assume that both Higgs fields develop real vacuum expectation values (VEVs) as $ \Phi_1 = \left( 0,  v_1\right) ^T, \Phi_2 =  \left(0 , v_2\right)^T $.
Then the electroweak scale, $v_{\mr EW}$ ($\simeq $ 246 GeV), can be expressed by these VEVs as $v_{\rm EW}^2 = 2v_1^2 + 2v_2^2$.

For later use, we introduce the Higgs fields in a two-by-two matrix form\cite{Grzadkowski:2010dj}, $H$, defined by $H = \left( i\sigma_2 \Phi_1^*,\ \Phi_2\right)$.
The field $H$ transforms under the electroweak $SU(2)_L \times U(1)_Y$ symmetry as $H  \to  \exp\left[\f{i}{2}\alpha_a(x) \sigma_a\right] H ~\exp\left[-\f{i}{2} \beta(x) \sigma_3\right]$, and therefore the covariant derivative on $H$ is expressed as $D_\mu H =\partial_\mu H - i \frac{g}{2} \sigma_a W_\mu^a H + i \frac{g'}{2}H\sigma_3 Y_\mu$.
The VEV of $H$ is expressed by a diagonal matrix $\langle H \rangle = \mr{diag} (v_1,v_2)$, and the potential can be written by using $H$ as follows:
\begin{align}
 & \hspace{-0.5cm}V(H)
 = - m_{1}^2~ \mr{Tr}|H|^2 - m_{2}^2~ \mr{Tr}\left(|H|^2 \sigma_3\right) - \left( m_{3}^2 \det H + \mr{h.c.}\right)\nonumber \\
& + \alpha_1~\mr{Tr}|H|^4  +  \alpha_2 ~\left(\mr{Tr}|H|^2 \right)^2+ \alpha_3~ \mr{ Tr}\left(|H|^2 \sigma_3 |H|^2\sigma_3\right)  \n \\
& + \alpha_4~ \mr{Tr}\left(|H|^2 \sigma_3 |H|^2\right)+ \left(\alpha_5 \det H^2 + \mr{h.c.}\right),
\end{align}
where $|H|^2 \equiv H ^\dagger H$ and we have imposed a (softly-broken) ${\mathbb Z}_2$ symmetry, $H \to H \sigma_3$ (or $\Phi_1 \to +\Phi_1$,  $\Phi_2 \to -\Phi_2$), in order to suppress Higgs-mediated flavor-changing neutral current processes.
In this paper, to make the discussion simpler, we take the five parameters $m_{2}, m_{3}, \alpha_3, \alpha_4,\alpha_5$ to 0.
The potential thus reduces to the following simple form: $ V(H) =  -m_1^2 \mr{Tr} |H|^2 + \alpha_1 \mr{Tr}|H|^4 + \alpha_2 \left(\mr{Tr}|H|^2 \right)^2 \label{010955_17Mar19}$.

The custodial transformation acting on the matrix $H$\cite{Grzadkowski:2010dj,Pomarol:1993mu} is defined as the following global $SU(2)$ transformation: $H \to U H U ^\dagger$, $U \in SU(2)_\mr{C}$ 
\footnote{
Note that this  $SU(2)_C$ transformation is different from the $U(2)$ basis transformation: $\Phi_i \to \sum_{j=1}^2M_{ij} \Phi_j$ ($i=1,2$).
}.
In addition, the $SU(2)_W$ gauge field also transforms as an adjoint representation.
The potential $V(H)$ has a symmetry under this transformation, which we call as the custodial symmetry.
However, the presence of the $U(1)_Y$ gauge field explicitly breaks the symmetry down to $U(1)_\mr{EM} \rtimes (\mathbb{Z}_2)_\mr{C} $, where $ (\mathbb{Z}_2)_\mr{C}$ transforms $H$ to $ i\sigma_1 H (i \sigma_1)^\dagger $  and ``$\rtimes$'' denotes the semidirect product because $ (\mathbb{Z}_2)_\mr{C} $ acts on $U(1)_\mr{EM}$.
Note that the symmetry $ (\mathbb{Z}_2)_\mr{C} $ follows from $m_2=\alpha_4=0$, and yields $\tan \beta \equiv v_2/v_1=1$.

In addition, since $m_3=\alpha_5=0$, the Lagrangian (not only the potential) is invariant under a global $U(1)_a$ transformation, which rotates the relative phase of the two doublets: $ H \to e ^ {i \alpha} H$ (or $\Phi_1 \to e^{-i\alpha} \Phi_1$, $\Phi_2 \to e^{i\alpha} \Phi_2$) ($0\leq\alpha < \pi$).
After $H$ gets a VEV, this $U(1)_a$ symmetry is spontaneously broken and the corresponding Nambu-Goldstone boson appears.\\

\section{Electroweak Strings}
In Refs.\cite{Eto:2018tnk,Eto:2018hhg,Dvali:1994qf,Dvali:1993sg}, it is pointed out that, unlike in the SM case, 2HDM allows topologically stable strings to exist thanks to the global $U (1)_a$ symmetry.
First, consider topological strings with the $Z$ flux (topological $Z$ strings).
There are two types of topological $Z$ strings corresponding to which one of the two Higgs doublets is to be wound.
To see that, let us take $W_\mu^\pm = A_\mu=0$.
Here we have defined as $Z_\mu \equiv W_\mu ^3 \cos\theta_W - Y_\mu \sin\theta_W$, $A_\mu\equiv W_\mu ^3 \sin\theta_W + Y_\mu \cos\theta_W$.

The solution called a $(1,0)$-string \footnote{$(1,0)$ means that the phase of $\Phi_1$ winds once around the circle at spatial infinity but that of $\Phi_2$ does not.} is given by
\begin{align}
 H^{(1,0)} &=v~ \mr{diag}\left( f(\rho) e^{i\varphi}, h  (\rho) \right),\\
 Z_i ^{(1,0)} &=  -\f{\cos \theta_W}{g} \f{ \epsilon_{3ij}x^j}{\rho^2} \left(1- w (\rho)\right)\label{012222_17Mar19},
\end{align}
where $v \equiv m/\sqrt{2\alpha_1 + 4 \alpha_2}~(=v_1=v_2)$, $\rho \equiv \sqrt {x ^ 2 + y ^ 2} $ and $ \varphi $ is the rotation angle around the $z$-axis.
The boundary conditions imposed on the profile functions are $f(0)={h'}(0)=w(\infty)=0$, $w(0)=f(\infty)=h(\infty)=1$.
Thus the asymptotic form of $H^{(1,0)}$ at infinity is $\sim v~\exp[{\f{i\varphi}{2}}]\exp[{\f{i\varphi}{2} \sigma_3}] $.
On the other hand, the solution called a $ (0,1) $-string is given by $ H^{(0,1)} = i\sigma_1  H^{(1,0)} ( i\sigma_1)^\dagger$ and $ Z_i ^{(0,1)} = - Z_i ^{(1,0)}$.
Both the $(1,0)$- and $(0,1)$-strings have winding number $1/2$ for the global $U(1)_a$ and thus they are topological vortex strings.
Note that both of them have logarithmically divergent tension due to the kinetic term of the Higgs field:
\begin{equation}
2\pi \int d\rho \rho ~\mr{tr}|D_i H^{(1,0)}|^2 \sim 2\pi \int d\rho \rho ~\mr{tr}|D_i H^{(0,1)}|^2 \sim \pi v^2 \int \f{d\rho} {\rho} \h{2em}(\mr{for}~\rho\to \infty)\label{202812_20Dec19},
\end{equation}
which is a quarter of that for a global $U(1)_a$ {\it integer vortex}
because of the half winding number for $U(1)_a$ \cite{Eto:2018tnk}.
On the other hand, the contribution from the half winding in the gauge orbit is
canceled by the gauge fields $Z_i^{(0,1)}$ or $Z_i^{(1,0)}$ and exponentially suppressed,
and as a consequence, the $Z$ flux is squeezed into the flux tube, 
which exponentially vanishes as a usual Abrikosov-Nielsen-Olesen vortex 
\cite{Abrikosov:1956sx,Nielsen:1973cs}.
%
In addition, the $Z$ fluxes flowing inside them are $ \pm 2 \pi \cos \theta_W / g $ along the $z$-axis, respectively, which are half of that of a non-topological $Z$ string in the SM.

Our potential has the custodial $SU (2)_\mr{C} $ symmetry.
When $\sin \theta_W=0 $, this symmetry is the actual symmetry of the Lagrangian.
However, the presence of the topological string solutions spontaneously break it down to $ U(1)_\mr{C}$, and thus they have $S^2$ ($\simeq SU(2)_\mr{C}/ U(1)_\mr{C}$) moduli.
Each point on the $S^2$ moduli space corresponds to a physically different string solution 
having a common winding number $1/2$ for the global $U(1)_a$.
We parametrize the $S^2$ moduli space by two parameters $0\leq \zeta \leq \pi$, $0\leq \chi<2\pi$, where $\zeta$ and $\chi$ correspond to the zenith and azimuth angles, respectively.
We identify the (1,0)-string, Eq.(\ref{012222_17Mar19}), as the one associated with the south pole of the $S^2$ moduli space, $\zeta = \pi$.
On the other hand, the (0,1)-string corresponds to the north pole,  $\zeta = 0$.
String solutions on a generic point of the $S^2$ moduli space can be obtained by acting an $SU(2)_\mr{C}$ transformation on the $(1,0)$-string.
In particular, we can construct a one-parameter family on $S^2$ connecting a $ (0,1) $-string (north pole) and a $ (1,0) $-string (south pole) 
using the $ \sigma^2 $-axis rotation $U(\zeta) = \mr{exp}({\f{i}{2}\zeta\sigma_2}) ~(0\leq \zeta \leq \pi)$,
which corresponds to the one-parameter path on $S^2$: $0\leq \zeta \leq \pi$, $\chi=0$.
This fact will be important in the construction of the magnetic monopole solution later.

\begin{figure}[t]
\begin{center}
\includegraphics[width=0.5\linewidth]{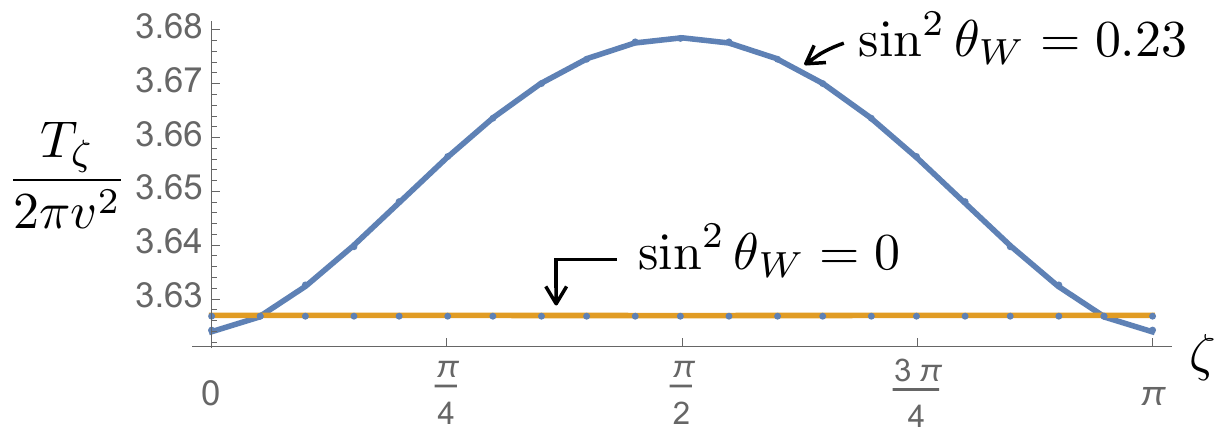}
\caption{Tension of the one-parameter family of the strings made with $U(\zeta)$.
$\zeta=0,\pi/2, \pi$ correspond to the north pole, the equatorial point and the south pole on the moduli space $S^2$, respectively.
Although this value itself depends on the infra-red cut off, the difference between the maximum and minimum values does not.}
\label{213442_4Apr19}
\end{center}
\end{figure}

We should note that in our case, $ \sin \theta_W \neq 0 $ and the custodial symmetry is not exact.
As a consequence, almost all the points of the $S^2$ moduli space are energetically lifted.
As studied in Refs.\cite{Eto:2018tnk,Eto:2018hhg}, the two $Z$ strings, $(1,0)$-string and $(0,1)$-string, 
are the most stable with degenerate among the topological strings.
On the other hand, solutions on the equatorial points of the moduli space, which contain a $W$ flux and are called as $W$ strings, are the most unstable.
Fig.\ref{213442_4Apr19} shows a plot of the string tension of the one-parameter family $U(\zeta) H^{(1,0)} U(\zeta) ^\dagger$.
As we stated above, the $ (\mathbb{Z}_2)_\mr{C} $ symmetry remains, 
which corresponds to flipping the upper and lower half spheres of the moduli space $ S^2$ (or $\zeta \to \pi-\zeta$) 
\footnote{The symmetry under rotations around the $ \sigma^3 $-axis also remains as $ U(1)_\mr{EM} $.}.
Thus, the presence of one of the $Z$ strings spontaneously breaks the $ (\mathbb{Z}_2)_\mr{C} $ symmetry.\\

\section{Magnetic monopole configuration}
Let us make a magnetic monopole configuration in 2HDM as follows.
First, prepare a (0,1)-string and a (1,0)-string, cut both in half and connect them smoothly.
In other words, make a configuration where the upper half ($z>0$) is a $(0,1)$-string, but the lower half ($z<0$) is a $(1,0)$-string.
Such a smooth connection is realized by the one-parameter path on the moduli space made with $U(\zeta)$ ($0 \leq \zeta \leq \pi$).
Since (0,1)-string and (1,0)-string have the $Z$ fluxes in the opposite directions, $Z$ fluxes of $ 4 \pi \cos \theta_W / g $ in total flows from the connection.
This is the same as the amount of a $Z$ flux that a Nambu monopole discharges.
Thus, the same argument for the conservation of a $U(1)_Y$ flux ensures that $ 4 \pi \sin \theta_W / g $ magnetic flux spreads from the connection, and thus it is a magnetic monopole.
Interestingly, this can be regarded as one in which the non-topological $Z$ string attached to a Nambu monopole is divided in two topological $Z$ strings and these two fractions are pulled to the opposite directions to each other.

Let us discuss the stability of such a configuration.
In the present case $ \tan \beta = 1 $ and thus the tensions of the two $Z$ strings are equal.
Therefore, the monopole will not be pulled to one side, and such configuration is expected to be stable.
This argument is intuitive and easy to understand, but we give a more rigorous discussion.
The stability is ensured by the following two reasons.
First, the whole of this configuration (both the strings and monopole) has a topological charge $1/2$ for the global $U(1)_a$ symmetry in any cross sections with $z=\mr{const.}$.
In other words, concentrating on the $U(1)_a$ charge, the configuration is an infinitely long global string.
Therefore, it cannot be broken into pieces as usual global strings, 
and thus $(1,0)$- and $(0,1)$-strings must always be connected by a path on the lifted moduli space $S^2$.
Second, the string tensions on $ S^2 $ has a degenerate double-well structure (Fig.\ref{213442_4Apr19}), where $(1,0)$- and $ (0,1)$-strings are the most stable, and thus the above connection can be regarded as a topological kink interpolating between the two minima of the $ (\mathbb {Z}_2)_\mr{C} $ symmetric potential.
Therefore, the stability of the monopole is topologically ensured as in the case of a $\mathbb{Z}_2 $ kink.
These arguments can be summarized in the statement that the stability is ensured by the $U(1)_a$ symmetry and the $(\mathbb{Z}_2)_\mr{C}$ symmetry.

We construct such a configuration concretely in the limit where we treat vortices as infinitesimally small (delta-function like) topological defects.
This limit corresponds to those far enough from vortices.
The actual regular solution will be constructed numerically by the relaxation later.

The Higgs matrix $ H^{(0,1)} $ of a (0,1)-string at large distance ($ \rho \to \infty $) is expressed as $H^{(0,1)}\to H_0 \equiv v ~\mr{diag}( 1, e^{i\varphi})$.
Let us introduce a function $\hat{\zeta}(r,\theta)$ and a unitary matrix $ U(\hat{\zeta}) = e^{\frac{i}{2}\hat{\zeta} \sigma_2}$, and then acting a ``local custodial transformation'' on $H_0$, we obtain $ H_{\hat{\zeta}} = U(\hat{\zeta}) H_0 U(\hat{\zeta})^\dagger$.
 Here, $r$ and $\theta$ are the distance from the origin and the zenith angle from the $z$-axis, respectively, 
and $\hat{\zeta}(r,\theta)$ is assumed to be a monotonically increasing function with respect to $ \theta $ and $\hat{\zeta}(r,0)=0,~ \hat{\zeta}(r,\pi)=\pi$.
We should note that $ H_{\hat{\zeta}} $ has a winding number $1/2$ for $ U(1)_a $ on any planes with $z=\mr{const.}$ 
and hence describes a global string along the $z$-axis.
In fact, its kinetic energy behaves as $ \mr{tr}|D_i H_{\hat{\zeta}}|^2\sim v^2/\rho^2$ for $\rho\to \infty$ 
and hence the tension on $z=\mr{const.}$ is logarithmically divergent independently of $\hat{\zeta}$ (or $z$),
where the tension from the winding in the gauge orbit is canceled by $\hat{\zeta}$-dependent gauge fields and remains finite 
(See Appendix~\ref{030312_19Apr19}).
Furthermore, the string behaves as (0,1)- and (1,0)-strings around the positive and negative sides of the $z$-axis ($ \theta \sim 0, \pi $), respectively,
while it behaves locally as the $W$-string in a region in which $\hat{\zeta} \sim \pi/2$
\footnote{The localized $W$-string carries a $U(1)$ modulus corresponding to points on the equator of the $S^2$ moduli space.}.
The two $Z$-strings connect at the origin, and thus the $Z$ flux flows upward and downward from the origin as we see later.
In addition, the electromagnetic flux emanates from the connection and it behaves as a magnetic monopole.

\section{$Z$ flux and magnetic flux}
The gauge fields induced from the Higgs field $H_{\hat{\zeta}}$ are determined to minimize the kinetic energy :$\int d^3x~ \mr{Tr}| D_i  H_{\hat{\zeta}} |^2$.
After minimizing, we obtain asymptotic forms of the field strengths of the $Z$ field and the electromagnetic field (See Appendix \ref{030312_19Apr19} for the derivation).
The former is given by 
\begin{align}
 F_{12}^Z &= \f{2\pi \cos \theta_W}{g}~\f{z}{|z|}~ \delta(x)\delta(y)\label{142119_14Mar19}
\end{align}
and  $F_{23}^Z = F_{31}^Z =0$.
From Eq.(\ref{142119_14Mar19}), it can be seen that $Z$ flux $2\pi \cos \theta_W/g$ flows on the $ z $-axis toward $ z = \pm \infty $ from the origin.

%
%
				   %
For the electromagnetic field, we obtain
\begin{equation}
 F_{ij}^{\mr{EM}} =  -\f{\sin \theta_W}{g}  \sin \hat{\zeta} ~ (\partial_{[i} \hat{\zeta})(\partial_{j]} \varphi) \label{154816_14Mar19},
\end{equation}
where $t_{[ij]} \equiv t _{ij} - t _{ji} $ for any tensor $t$.
From Eq.~(\ref {154816_14Mar19}), it is clear that there is a magnetic flux in a region where $\partial_{i}  \hat{\zeta} \neq 0$, and that it is coming out of the connection at the origin.
The total magnetic flux $ \Phi _B$ can be obtained by integrating the flux density $ B_i \equiv -\f{1}{2} \epsilon_{ijk} F_ {jk} ^ {EM} $ on an infinitely large sphere covering the system and using the Gauss's theorem:
\begin{equation}
 \Phi _B = \f{2\pi \sin\theta_W}{g} \int _{-\infty}^\infty dz ~ \partial_3 \cos \hat{\zeta} =  \f{4\pi\sin \theta_W}{g}, 
\end{equation}
where we have used $\hat{\zeta}(r,0)=0$ and $ \hat{\zeta}(r,\pi)=\pi$ in the last equality.
Therefore, there is a magnetic monopole at the origin with a magnetic charge $ 4 \pi \sin \theta_W / g $.
This monopole carries a $U(1)$ modulus coming from that of the localized $W$-string.

Interestingly, the above analysis does not rely on details in the form of $ \hat {\zeta} (r, \theta) $.
The existence of magnetic monopole is determined only by the information of the end points that $ \hat{\zeta} = 0, \pi $ for $ \theta = 0, \pi $, respectively.
This property is similar to that of topological kinks.
On the other hand, the ``shape'' of the magnetic flux spreading from the monopole depends on the functional form of $ \hat {\zeta} (r, \theta)$.
By minimizing the magnetic energy, it is determined to be $\hat{\zeta} = \theta$ and we obtain the magnetic flux density $B_i = (\sin \theta_W/g) ~x_i/r^3$, which means that the magnetic flux is distributed spherically  at large distances from the monopole.\\
 
\section{Numerical simulation}
Here we show a stable regular magnetic monopole solution numerically constructed by relaxation.
The procedure is as follows.
We smear out the infinitely small defects (singularities) in the configuration discussed above using some profile functions with a typical scale $\sim v^{-1}$.
The smeared configuration is regular anywhere and approaches at large distances to the asymptotic form constructed above.
Then, we evolve it by the relaxation until it sufficiently approaches the solution of the equation of motion. 
We have taken the parameters $g,g',m_1,\alpha_1,\alpha_2$ so that the physical parameters are given by
$\sin ^2 \theta_W  = 0.23, ~ m_W = 80~ \mr{GeV}, ~ v_{\mr{EW}} = 246~ \mr{GeV},~ m_h = 125~ \mr{GeV},~ m_{H} = m_{H^\pm} = 400 ~\mr{GeV}$,
where $m_W$ and $m_h$ are masses of the $W$ boson and the SM Higgs, respectively.
In addition, $m_{H}$ and $m_{H^\pm}$ are the masses of the heavier CP-even Higgs and the charged Higgs bosons, respectively.

To carry out the numerical computation, we have used a length unit in which $v$ is normalized to unity.
We have computed the relaxation in a 3D box with the size $L_x=L_y=8$, $L_z=12$ and adopted the Dirichlet boundary conditions
such that the field values on the boundaries are fixed to the asymptotic ones constructed above.
After the relaxation time $t = 20$, the variation of the energy density per time is $\mathcal{O}(10^{-5})$, 
thus we have regarded the convergence achieved.
We also have confirmed that the field values converged up to the same order.

\begin{figure*}[t]
\begin{center}
\includegraphics[width=1.0\textwidth]{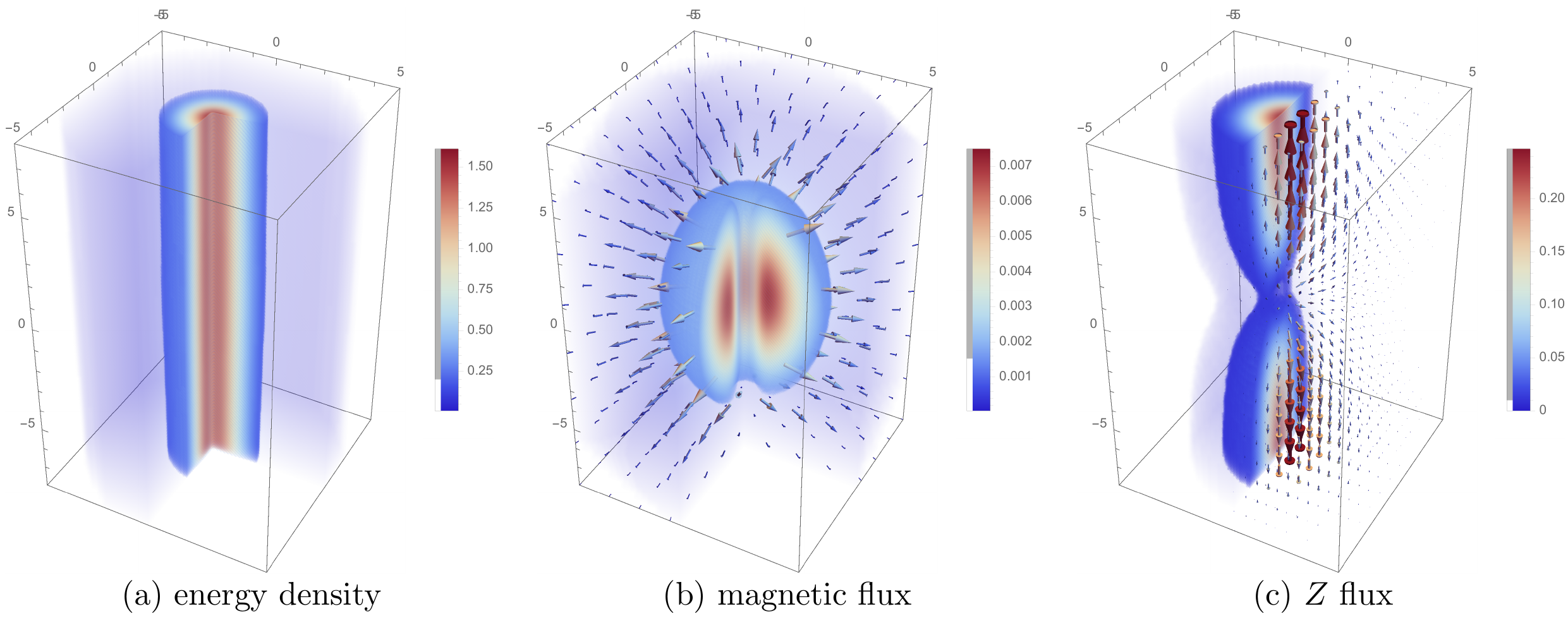}
\caption{Plots for the numerical solution constructed by the relaxation. 
In all plots, $v$ is normalized to unity.
(a): Energy density. 
The color represents its value, where red is the largest and blue is the smallest. 
There is a string-like object that contains the energy density along the $z$ axis.
(b): Magnetic flux. 
The direction of the arrows indicates that of the flux. 
Also, the color and size of the arrows indicate the flux density, where red is the strongest, blue is the weakest.
We can see the existence of a magnetic monopole at the center from which the magnetic flux emanates.
The flux is spherical at large distances, but not in the vicinity of the monopole.
(c): $Z$ flux. 
The direction, color and size of the arrows are the same as those for the magnetic flux. 
The $Z$ flux flows upward and downward along the string from the monopole.}
\label{155536_4Apr19}
\end{center}
\end{figure*}

Fig.\ref{155536_4Apr19} shows plots of the energy density, magnetic flux and $Z$ flux of the solution.
The energy is localized in the form of a string.
Also, the magnetic flux is rising from the center, and it is clear that the magnetic monopole exists.
Note that the flux is spherically symmetric at large distances but not near the monopole.
The $Z$ flux flows upward and downward on the string, which indicates that the two $Z$ strings are attached to the monopole.

We would like to emphasize that
the energy density at any cross sections with $z=\mr{const.}$ includes the power-law tail leading to the logarithmic divergence,
but it is only related to the global $U(1)_a$ winding.
Therefore, neither the $Z$-fluxes at the both sides of the configuration,
which decay exponentially fast, nor the electromagnetic monopole suffer from the
logarithmic divergence.
We define the total energy of the monopole (monopole mass) as
the difference between the energy of the configuration and that of the $Z$-string without the monopole,
which has no divergence and is well-defined
because the energy density of the monopole (the difference of their energy densities) is exponentially localized.
In our numerical computation, the monopole mass is typically $\mathcal{O}(1)~\mr{ TeV}$.

\section{Conclusion and outlook}
We have constructed a stable magnetic monopole solution in 2HDM under a potential with the $U (1) _a $ symmetry and the $(\mathbb{Z}_2)_\mr{C}$ symmetry, which yields $\tan \beta =1$, based on full 3D numerical simulation.
In the solution, the magnetic flux $4\pi\sin\theta_W/g$ emanates from the monopole, and the $Z$ strings are attached to its both sides.
The stability is ensured by the fact that the monopole can be regarded as a topological $(\mathbb{Z}_2)_\mr{C}$ kink interpolating between the two degenerate vacua on the lifted moduli space $S^2$.

We should note that our monopole is essentially different from monopoles confined in vortices, which are put in the Higgs phase \cite{Tong:2003pz,Eto:2006pg,Auzzi:2003em,Shifman:2004dr,Hanany:2004ea,Eto:2004rz,Tong:2005un,Shifman:2007ce,Shifman:2009zz}
\footnote{
In particular, confined monopoles in dense QCD \cite{Gorsky:2011hd,Eto:2011mk} are quite similar to our monopole 
in the sense that they are accompanied by both the color flux tubes and global vortices,
although they have no magnetic flux spreading spherically from them.
}.
In those cases, the magnetic flux and the physical degrees of freedom are confined in the string.
On the other hand, our monopole is not in the Higgs phase since $U(1)_\mr{EM}$ is not broken.
Consequently, while the $Z$ flux is confined, the magnetic flux is not confined but emanates spherically at large distances.

In this paper, we have assumed the $U(1)_a$ symmetry, and thus the corresponding NG boson appears after $H$ gets a VEV.
Since such massless modes are severely constrained by experiments, we should introduce explicit $ U(1) _a $ breaking terms by switching on $ m_{3}$ and $ \alpha_5 $, which give a mass to the NG boson (the CP-odd Higgs boson).
This effect makes the topological strings attached by domain walls \cite{Eto:2018tnk,Eto:2018hhg}.
The same is expected to happen with the string-monopole complex that we have considered above.
As a result, our monopole (and string) are pulled by the wall and cannot be static.
There is another option to avoid the NG boson, which is gauging the $ U (1) _a $ symmetry \cite{Ko:2012hd}.
In such models, the monopoles are not attached by domain walls because the $U(1)_a $ symmetry is exact.
Therefore, their evolution is easier to study.

Let us comment on phenomenological properties of our monopole.
Interestingly, they did not be diluted by the cosmological inflation because it is produced at the electroweak symmetry breaking.
This raises a question of whether our monopole causes the so-called cosmological monopole problem \cite{Preskill:1979zi}.
A naive answer is no, because our monopole has a much lighter mass $\mathcal {O}(1) ~ \mr{TeV}$ than that of GUT monopoles,
and thus it would not dominate the energy density of the universe at any epoch assuming that a few monopoles per horizon volume were produced at the symmetry breaking.
Another famous bound on monopoles is the Parker bound \cite{Parker:1970xv,Turner:1982ag}.
Again, our monopole would not conflict the bound because of its lightness.
Of course, these arguments are quite naive.
Therefore, it would be interesting to investigate the evolution of our monopoles (and strings) in the early universe and the abundance in the present universe.
Such studies provide predictions for the current and nearly future monopole searches, and enable us to impose bounds on 2HDM.
To this end, it will also be important to consider whether the monopoles can be generated, not only by the Kibble-Zurek mechanism \cite{Kibble:1976sj,Zurek:1985qw}, but also by the reconnection between the $Z$ strings.

%

Another phenomenological application of our result may be baryogenesis.
In Ref.~\cite{Vachaspati:1994ng}, a sphaleron-like configuration is made from a Nambu monopole-antimonopole pair (dumbbell) in the SM.
It is interesting to find out whether the same argument is applicable to our monopole.
In addition, it is also interesting to consider the direct detection of the monopoles by collider experiments.

Finally, let us briefly comment on the case without the $(\mathbb{Z}_2)_\mr{C}$ symmetry, in which the $Z$ fluxes of the two $Z$ strings are not equal, and a difference occurs in the finite parts of those tensions.
Therefore, the monopole is pulled to one side, resulting in the instability.
A quantitative study of the instability will be done elsewhere.
%

\section*{Acknowledgements} 
We would like to thank Nobuyuki Matsumoto for useful discussions.
%
This work is supported by 
the Ministry of Education, Culture, Sports, Science (MEXT)-Supported Program for the Strategic Research Foundation at Private Universities ``Topological Science” (Grant No. S1511006)''. 
The work is also supported in part by JSPS Grant-in-Aid for Scientific Research 
(KAKENHI Grant No. JP16H03984 (M.~E. and M.~N.),
 No. JP19K03839 (M.~E.),
 No. JP18J22733 (Y.~H.),
 No. JP18K03655 (M.~K.),
 No. JP18H01217 (M.~N.)), 
and also by MEXT KAKENHI Grant-in-Aid for Scientific Research on Innovative Areas “Topological Materials Science” No. JP15H05855 (M.~N.) 
and “Discrete Geometric Analysis for Materials Design” No. JP17H06462 (M.~E.)
 from the MEXT of Japan.

\appendix

\section{Gauge field induced by $H$}
\label{030312_19Apr19}
In the presence of the Higgs configuration $H_{\hat{\zeta}}$, the electromagnetic flux and the $Z$ flux are induced.
In this section, we derive them in the limit where we treat the strings and the monopole as infinitesimally small (delta-function like) defects.

To make it easy to see the correspondence with a Nambu monopole, in the following, we consider an alternative Higgs field, 
\begin{align}
 H_{\mr{mon}}&= U(\hat{\zeta})^\dagger H_{\hat{\zeta}} = v
       \begin{pmatrix}
  \cos \f{\hat{\zeta}}{2} & - \sin \f{\hat{\zeta}}{2} \\
   e^{i\varphi} \sin \f{\hat{\zeta}}{2} &  e^{i\varphi} \cos\f{\hat{\zeta}}{2}
	\end{pmatrix} ,
\end{align}
which is $SU(2)_W$ gauge equivalent to $H_{\hat{\zeta}} $.\\

Let us minimize the kinetic energy of $H_{\rm{mon}}$:
\begin{equation}
 \int d^3x~ \mr{Tr}| D_i H_{\rm{mon}} |^2 = \int d^3x\left(| D_i \Phi_{1} |^2 +  | D_i \Phi_{2} |^2 \right).\label{210854_20Dec19}
\end{equation}
If $ H_{\rm{mon}} $ would be a local string, we can take the gauge field to satisfy $ D_i \Phi_1 = D_i \Phi_2 = 0 $, but now since $ H_ {\mr{mon}} $ has a global winding $1/2$, its kinetic energy cannot be completely canceled by the gauge fields.
Therefore, it is minimized by the following:
\begin{equation}
  D_i \Phi_{1} = - \f{i}{2} (\partial_i \varphi) \Phi_{1},\label{112302_14Mar19} 
\end{equation}
\begin{equation}
 ~ D_i \Phi_{2} = + \f{i}{2} (\partial_i \varphi) \Phi_{2}\label{112312_14Mar19}.
\end{equation}
Since $ \Phi_{1}$ and $ \Phi_{2}$ are given by
\begin{equation}
  \Phi_{1} = v e^{-i\f{\varphi}{2}}
  \begin{pmatrix}
   -\sin \f{\hat{\zeta}}{2} e^{-i \f{\varphi}{2}}\\
   \cos \f{\hat{\zeta}}{2}e^{i\f{\varphi}{2}}		   
\end{pmatrix},
\end{equation} 
\begin{equation}
 \Phi_{2} = v e^{i\f{\varphi}{2}}
  \begin{pmatrix}
   -\sin \f{\hat{\zeta}}{2} e^{-i\f{\varphi}{2}}\\
   \cos \f{\hat{\zeta}}{2}e^{i\f{\varphi}{2}}		   
  \end{pmatrix},
\end{equation}
Eqs.(\ref{112302_14Mar19}) and (\ref{112312_14Mar19}) are equivalent.
Thus, it is sufficient to consider only Eq.(\ref{112302_14Mar19}).
Following the procedure in Ref.\cite{Nambu:1977ag}, we obtain the gauge fields 
\begin{align}
 g W_i^a &= - (\overrightarrow{n}\times \partial_i \overrightarrow{n})^a + h_i n^a ,\label{120157_14Mar19}\\
 g' Y_i &= a_i (r,\theta,\varphi)\label{120204_14Mar19},
\end{align}
with
\begin{equation}
 h_i \equiv \partial_i \varphi - i v^{-2} (\Phi_1 ^\dagger~ \overset{\leftrightarrow}{\partial_i}~ \Phi_1) -a_i.
\end{equation}
Here, $ a_i (r, \theta, \varphi) $ is an arbitrary function at present, and $n^a$ is a unit vector defined by $ n^a \equiv~ (\Phi_1^\dagger \sigma^a \Phi_1 + \Phi_2^\dagger \sigma^a \Phi_2)/(\Phi_1^\dagger \Phi_1+\Phi_2^\dagger \Phi_2)$, which indicates the orientation of the vacuum within $SU(2)_W$.
It has the hedgehog structure and plays the same role as the normalized adjoint scalar in the presence of a 't Hooft-Polyakov monopole \cite{tHooft:1974kcl,Polyakov:1974ek}.

Using Eqs.(\ref{120157_14Mar19}) and (\ref{120204_14Mar19}), we can obtain the field strengths, $W_{ij}^a$ and $Y_{ij}$, as follows:
\begin{align}
  g n^a W_{ij}^a &= \left(2 \sin^2 \f{\hat{\zeta}}{2}-1\right) \partial_{[i} \partial_{j]} \varphi  - \partial_{[i}a_{j]}\label{001852_11Apr19} \\
g' Y_{ij} &= \partial_{[i} a_{j]}.\label{014950_20May19}
\end{align}
In our notation, $t_{[ij]} \equiv t _{ij} - t _{ji} $ for any tensor $t$.

Let us define the field strengths of the $Z$ field and the electromagnetic field as follows
\footnote{These definitions might not be valid when $ (D_i n) ^ a \neq 0 $, that is, in the vicinity of strings or monopoles. However, that is enough because we are only interested in the behavior at large distances. For related arguments, see Refs.\cite{Tornkvist:1997hd,Tornkvist:1998sm}.}
\begin{align}
 F_{ij}^Z &\equiv - \cos \theta_W n^a W_{ij} ^a - \sin \theta_W Y_{ij}\label{131635_29Mar19}, \\
 F_{ij} ^{\mr{EM}} &\equiv -\sin \theta_W n^a W_{ij} ^a + \cos \theta_W Y_{ij}.
\end{align}
Substituting Eqs.(\ref{001852_11Apr19}) and \ref{014950_20May19} into Eq.(\ref{131635_29Mar19}), we obtain 
\begin{align}
 F_{12}^Z &= \f{2\pi \cos \theta_W}{g}~\f{z}{|z|}~ \delta(x)\delta(y)
\end{align}
and  $F_{23}^Z = F_{31}^Z =0$, where $a_i$-dependence has been canceled. 
On the other hand, $F^\mr{{EM}}_{ij}$ depends on $a_i$:
\begin{align}
 F_{ij}^{EM} &= - \f{\sin \theta_W}{g} \left(1-2 \sin^2 \f{\hat{\zeta}}{2}\right) \partial_{[i} \partial_{j]} \varphi +\f{\partial_{[i}a_{j]}}{g\sin \theta_W} .
\end{align}
To determine $a_i$, we require it to minimize the electromagnetic energy $ \int d^3x ~ (F_{ij} ^\mr{{EM}})^2 $, which leads to
\begin{equation}
 a_i = \sin ^2 \theta_W \left(1-2\sin^2 \f{\hat{\zeta}}{2}\right) \partial_i \varphi + b ~\partial_i \hat{\zeta},\label{020445_20May19}
\end{equation}
with $b$ being a constant.
The second term in Eq.(\ref{020445_20May19}) does not appear in the field strength, and we obtain the following expression:
\begin{equation}
 F_{ij}^{\mr{EM}} =  -\f{\sin \theta_W}{g}  \sin \hat{\zeta} ~ (\partial_{[i} \hat{\zeta})(\partial_{j]} \varphi) .
\end{equation}

\bibliographystyle{apsrev4-1}
\bibliography{./references}

\end{document}